# High-resolution Observations of a C9.3 White-light Flare and Its Impact on the Solar Photosphere

Zhe Xu[1,2], Xiaoli Yan[1,2], Zhentong Li[3], Liheng Yang[1,2], Zhike Xue[1,2], Jincheng Wang[1,2], and Yian Zhou[1,2]
[1] Yunnan Observatories, Chinese Academy of Sciences, 396 Yangfangwang, Guandu District, Kunming 650216, People's Republic of China; xuzhe6249@ynao.ac.cn
[2] Yunnan Key Laboratory of Solar Physics and Space Science, 396 Yangfangwang, Guandu District, Kunming 650216, People's Republic of China
[3] Purple Mountain Observatory, Chinese Academy of Sciences, No.8 Yuanhua Road, Qixia District, Nanjing 210034, People's Republic of China
*Received 2025 January 26; revised 2025 May 23; accepted 2025 May 26; published 2025 June 9*

## Abstract

We present a detailed analysis of a C9.3 white-light flare using high-resolution observations from the New Vacuum Solar Telescope. The flare occurred near the eastern solar limb on 2023 September 11, within NOAA AR 13431, and produced beam electrons with energies just below 50 keV as observed by the Hard X-ray Imager onboard the Advanced Space-based Solar Observatory. Two white-light flare kernels were detected in the TiO band, connected by filamentary brightenings aligned with penumbral fibrils, suggesting a photospheric contribution to the white-light emission. Notably, the impact of the flare on the solar photosphere was characterized by sudden vortex flows and significant amplification of the magnetic field in the white-light flare kernel region. We infer that this impact is driven by the propagation of flare-generated Alfvén wave pulses, which deposited energy into the photosphere. These observations support the potential role of the Alfvén wave mechanism in driving energy transport and heating during white-light flares.

*Unified Astronomy Thesaurus concepts:* Solar active regions (1974); Solar flares (1496); Solar activity (1475)

*Materials only available in the* online version of record: animation

## 1. Introduction

Solar white-light flares are observationally defined as the sudden enhancement in the visible optical continuum, appearing as intense, localized bright regions on the solar surface (D. F. Neidig 1989; H. S. Hudson 2011; C. Fang et al. 2013; H. S. Hudson 2016). Observational and statistical studies have shown that major flares, such as X- and M-class flares, are more likely to be identified as white-light flares, suggesting that the amount of energy released is a key factor in producing white-light emission (Y. Xu et al. 2006; Q. Hao et al. 2017; Y. Song & H. Tian 2018; Y. Li et al. 2024b, 2024c; Z. Jing et al. 2024). Although most sub-M-class flares typically do not show detectable white-light signatures, some C-class flares have nevertheless been observed to produce noticeable white-light emission (H. S. Hudson et al. 2006; D. B. Jess et al. 2008; Y. L. Song et al. 2018; Y. Song et al. 2020; Q. Li et al. 2024a; D.-C. Song et al. 2025). Y. Song & H. Tian (2018) reported that about 8% of sub-M-class flares exhibit white-light enhancements, and interestingly, the occurrence of such emission appears to be independent of the total energy released. A recent statistical study significantly increased the occurrence rate of white-light flares among C-class flares, reaching up to 30%, and revealed that in each level, the confined flares exhibit a higher occurrence of white-light flares compared to the eruptive flares (Y. Cai et al. 2024a). These results suggest that additional factors may also influence the production of white-light emission, such as magnetic field complexity (Y. Song & H. Tian 2018) and the height of the flare magnetic reconnection (M. D. Ding et al. 1999;

P.-F. Chen et al. 2001; Q. Hao et al. 2012; Y. Song et al. 2020). Distinguished from ordinary flares, white-light flares are capable of depositing a significant amount of energy into the deeper layers of the solar atmosphere, such as the lower chromosphere and photosphere, making them particularly valuable for studying the mechanisms of energy transport and heating across different atmospheric layers during solar flares.

It is widely accepted that nonthermal electron beam bombardment is a primary mechanism for the production of white-light flares (H. S. Hudson 1972; J. Aboudarham & J. C. Henoux 1987; D. F. Neidig 1989; L. Fletcher et al. 2007), as most white-light flares are temporally and spatially correlated with hard X-ray sources (K. Watanabe et al. 2010; S. Krucker et al. 2011; D. Li et al. 2023; Z. Jing et al. 2024). However, the precise formation height of white-light emission remains under debate, regarding whether the origin is photospheric (J.-C. Martínez Oliveros et al. 2012; V. Yurchyshyn et al. 2017), chromospheric (S. Krucker et al. 2015; M. Penn et al. 2016), or in between (G. S. Kerr & L. Fletcher 2014; L. Kleint et al. 2016; J. Hong et al. 2018). In the context of the photospheric origin, electron beams alone may not be sufficient to explain the observed heating, since nonthermal electrons typically possess energies below 100 keV (M. Kuhar et al. 2016; Q. Hao et al. 2017; Z. Jing et al. 2024), insufficient for direct penetration to such depths (V. Yurchyshyn et al. 2017). To address this limitation, mechanisms such as radiative backwarming (M. E. Machado et al. 1989; M. D. Ding et al. 2003; T. R. Metcalf et al. 2003), chromospheric condensation (W. Q. Gan et al. 1992; W. Q. Gan & P. J. D. Mauas 1994), and Alfvén wave dissipation (A. G. Emslie & P. A. Sturrock 1982; L. Fletcher & H. S. Hudson 2008; J. W. Reep & A. J. B. Russell 2016) have been proposed. It is also worth noting that Alfvén waves are frequently highlighted as a potential mechanism for both heating and electron reacceleration across different atmospheric layers (A. G. Emslie & P. A. Sturrock 1982; L. Fletcher & H. S. Hudson 2008; J. C. Brown et al. 2009; M. Varady et al. 2014; J. W. Reep & A. J. B. Russell 2016), particularly when considering possible deep layer origins for white-light emission (M. Varady et al. 2014; S. Krucker et al. 2015; D.-C. Song et al. 2023).

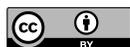

A key clue to the existence of the Alfvén wave pulse lies in the frequent association of white-light flares with sudden changes in the photospheric magnetic field during the impulsive phase (Z. Xu et al. 2016; Y. L. Song et al. 2018; J. S. Castellanos Durán & L. Kleint 2020; F. Zuccarello et al. 2020; L. Gong et al. 2024), even in a small C-class white-light flare (Q. Li et al. 2024a). Statistical studies have shown that the majority of white-light flares are accompanied by changes in the line-of-sight magnetic field (Y. L. Song et al. 2018; J. S. Castellanos Durán & L. Kleint 2020), with stronger white-light emissions correlating with greater magnetic field changes (Y. L. Song et al. 2018). These abrupt changes in the photospheric magnetic field suggest that energy stored in the coronal magnetic field is rapidly transmitted to the photosphere during the flare (E. W. Cliver et al. 2012; Y. Bi et al. 2018). However, the causal relationship between the white-light emission and photospheric magnetic field changes remains poorly understood. L. Fletcher & H. S. Hudson (2008) proposed that the impulsive reconfiguration of the coronal magnetic field could generate MHD fast-mode and Alfvén-mode wave pulses during flares (see Figure 1 in L. Fletcher & H. S. Hudson 2008). Fast-mode wave pulses, behaving as shrinking loops, can explain the changes in the line-of-sight photospheric magnetic field and the magnetic inclination angle around the flaring polarity inversion line (PIL; H. Wang & C. Liu 2010; Z. Xu et al. 2016, 2017; M. S. Wheatland et al. 2018; H. Chen et al. 2024; L. Gong et al. 2024). In contrast, Alfvén-mode wave pulses, which manifest as the propagating twists, can reaccelerate electrons, enabling them to penetrate deeper into the solar atmosphere and heat local plasma, thereby producing white-light emission. Alfvén-mode wave pulses are also thought to directly induce magnetic field changes at the footpoints, as they propagate along magnetic loops (L. Fletcher & H. S. Hudson 2008). However, clear observational evidence linking large-scale Alfvén wave pulses to energy transport to the photosphere, as well as the associated magnetic field changes and white-light emission, remains lacking.

In this study, we presented a detailed analysis of a C9.3 white-light flare and its impact on the solar photosphere. High-resolution observations from the New Vacuum Solar Telescope (NVST; Z. Liu et al. 2014; X. L. Yan et al. 2020) and the Solar Dynamics Observatory (SDO; W. D. Pesnell et al. 2012) provided clear insights into the white-light emission, photospheric plasma motions, and magnetic field changes during this flare. Additionally, data from the Hard X-ray Imager (HXI; Z. Zhang et al. 2019; Y. Su et al. 2022) onboard the Advanced Space-based Solar Observatory (ASO-S; W. Gan et al. 2023) and spectral observations from the Chinese Hα Solar Explorer (CHASE; C. Li et al. 2022) offer a comprehensive view of this event. These combined observations provide a unique opportunity to explore flare-induced Alfvén wave pulses and their potential connection to white-light flares.

## 2. Instruments and Data Reduction

NVST provides high-resolution observations of the chromosphere and photosphere through its Hα and TiO bands, respectively. The Hα channel observes the Hα line center (6562.8 Å) and its two off-bands (±0.6 Å), with a bandwidth of 0.25 Å, a field of view (FOV) of $180'' \times 180''$, a pixel size of $0\overset{''}{.}165$, and a cadence of 44 s. The TiO channel observes the TiO molecular band and its nearby continuum, which centered at 6562.8 Å with a bandwidth of 10 Å. The TiO channel has a smaller FOV of $100'' \times 80''$, with a pixel size of $0\overset{''}{.}052$ and a cadence of 30 s. NVST observations captured the target flare over the time period of 05:10–09:28 UT on 2023 September 11. The images are mapped to the heliographic coordinate by registering them to SDO observations using an automatic mapping technique (K. Ji et al. 2019; Y. Cai et al. 2024b).

The Atmospheric Imaging Assembly (AIA; J. R. Lemen et al. 2012) on board SDO provides extreme-ultraviolet (EUV) and ultraviolet (UV) images with a pixel size of $0\overset{''}{.}6$. The cadences are 12 s for EUV bands and 24 s for UV bands. The Helioseismic and Magnetic Imager (HMI; J. Schou et al. 2012), also on board SDO, provides photospheric continuum intensity images, line-of-sight (LOS) magnetograms, and vector magnetograms via observing the Fe I 6173 Å with a pixel size of $0\overset{''}{.}5$. The continuum intensity images and the LOS magnetograms have a cadence of 45 s, while the vector magnetograms, available as Space-weather HMI Active Region Patches (SHARP), are adopted with a cadence of 720 s.

CHASE provides the full-disk spectral observations of the Hα line (6559.7–6565.9 Å) and the Fe I line (6567.8–6570.6 Å) using a raster scanning mode. It achieves spectral, spatial, and temporal resolutions of 0.05 Å pixel$^{-1}$, ∼1″, and ∼73 s, respectively. The calibrated CHASE L1.5 data are adopted (Y. Qiu et al. 2022). The CHASE Fe I continuum here used is defined by a specific wavelength at 6568.6 Å, as suggested by Q. Li et al. (2024a) following a spectral line fitting.

ASO-S/HXI is a hard X-ray imaging spectrometer observing the Sun in an energy range of 10–300 keV, with a time cadence automatically adjusted according to the total X-ray flux between 0.125 and 4 s (Z. Zhang et al. 2019). The X-ray image of ASO-S/HXI was reconstructed by CLEAN (in HXI GUI v1.42 beta) for the energy range of 14–20 keV during 06:00:46 to 06:01:20 UT, corresponding to the white-light flare peak phase. The spatial resolution of the reconstructed HXI X-ray image is $6\overset{''}{.}5$. We use the imaging detectors from D19 to D91, all of which have been calibrated (Y. Su et al. 2024). The orbit counts 48 hr before are selected as background signals for imaging as well as for the following spectral analysis (Z. Li et al. 2025). The reconstructed X-ray image is then coaligned with the NVST Hα image at 06:01 UT, given that it has a similar distribution to the chromospheric brightenings.

The photospheric flow fields were calculated using the dense optical flow method in the OpenCV package (cv2.calcOpticalFlowFarneback), which is based on Gunnar Farnebäck's algorithm (G. Farnebäck 2003). Optical flow refers to the apparent motion of brightness patterns between consecutive images, and the motions are tracked based on assumptions of brightness constancy, spatial coherence, and small motion. Several parameters (pyr_scale = 0.5, levels = 5, iterations = 10, poly_n = 5, and poly_sigma = 1.2) were set according to the default recommendations and were not sensitive in our measurements. The method was then applied to NVST TiO images to derive the proper motion ($v_x$ and $v_y$) on the solar surface. The 11 pixel window (winsize = 11), corresponding to a spatial scale of approximately 0″.6, was determined empirically to provide stable results while maintaining sufficient spatial detail. The flow fields were computed for two time periods 05:54:45−05:59:45 UT and 06:05:15–06:10:15 UT, representing the photospheric velocity fields immediately before and after the flare. Each time period consisted of 11 TiO images, recorded under relatively stable atmosphere seeing conditions. The uncertainty in the flow field is approximately 10%, which were assessed using the endpoint error test.

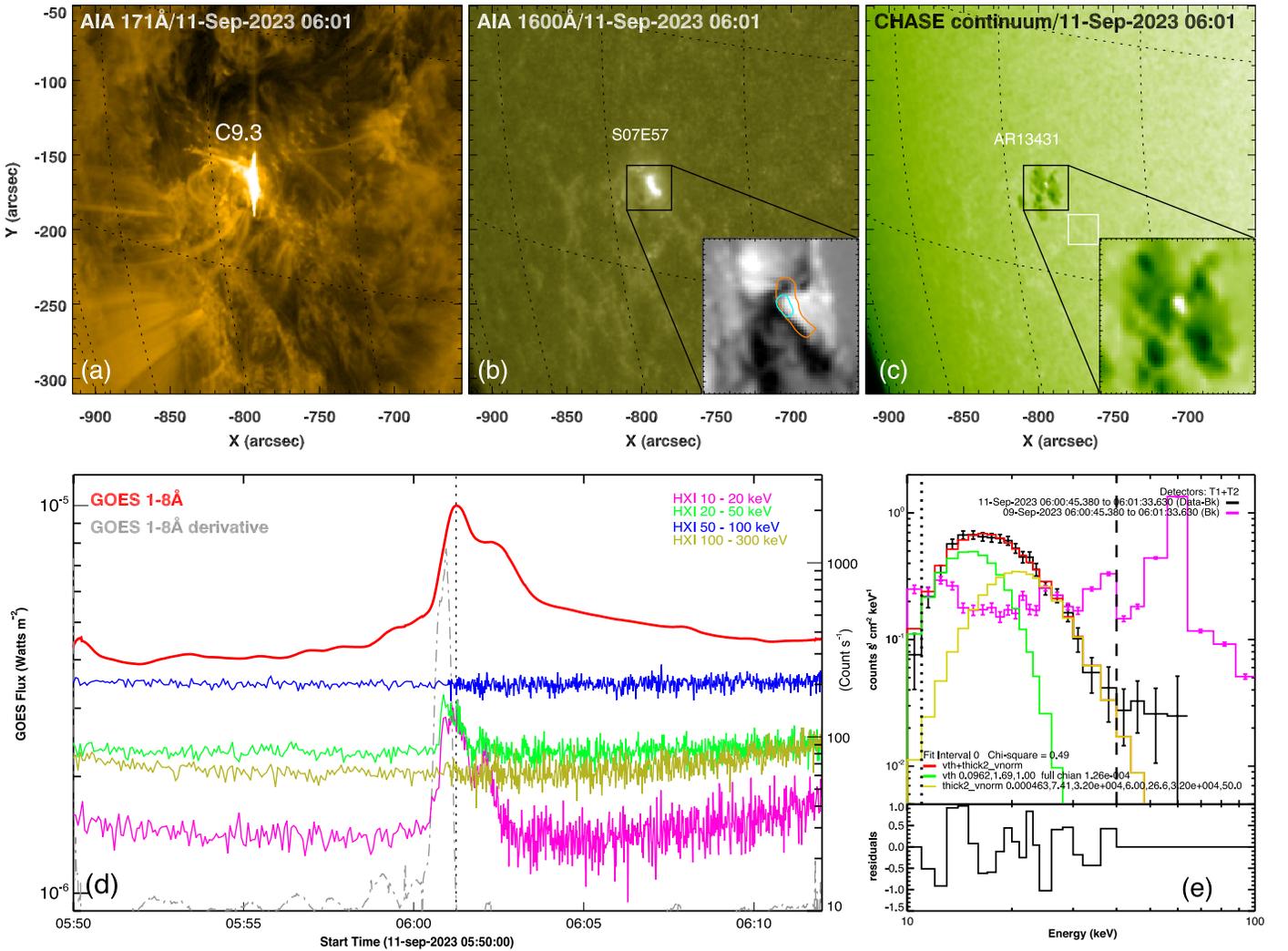

**Figure 1.** Panels (a)–(c): AIA 171 Å, 1600 Å, and CHASE continuum images showing the flare SOL2023-09-11T06:01 in AR 13431. The insets in panels (b) and (c) highlight the flare within the active region characterized by a δ-type magnetic configuration. The orange and cyan contours outline the flare brightenings observed in the AIA 1600 Å and CHASE continuum images, respectively. The white box in panel (c) marks the quiet-Sun region, which is used in Figure 2 to normalize the photospheric radiation, as well as to calculate the uncertainties in each light curve. Panel (d): light curves of GOES soft X-ray 1–8 Å flux, its time derivative, and the ASO-S/HXI HXR fluxes corresponding to the white-light flare. Panel (e): HXI HXR background-subtracted count spectrum (black curve) fitted using a combination of a thermal (green curve) and a nonthermal thick target model (yellow curve) at the peak time of the HXR emission. Two vertical dashed lines mark the energy range (11–40 keV) used for the spectral fitting.

### 3. Results

#### 3.1. Overview of the C9.3 White-light Flare

The target flare SOL2023-09-11T06:01 (C9.3) occurred in NOAA active region 13431 (S07E57) on 2023 September 11.

Figures 1(a)–(c) provide an overview of the flare in the corona, chromosphere, and photosphere around its peak time. This flare was accompanied by a solar jet, which was concentrated in a compact region as shown in the AIA 171 Å image. In the chromosphere, the flare produced a loop-shaped brightening, evident in the AIA 1600 Å image. In the photosphere, it generated a prominent brightening point, clearly visible in the CHASE continuum image, confirming that this event was a typical white-light flare. The magnetogram in the subpanel of Figure 1(b) reveals that the flaring region exhibited a magnetic δ configuration, which is a pattern of flare productivity. Notably, both the chromospheric loop-shaped brightening and the photospheric brightening point were located near the magnetic PIL of the δ-type sunspot, as illustrated by the orange and cyan contours on the magnetogram.

Figure 1(d) shows the light curves of GOES soft X-ray (SXR) and ASO-S/HXI Hard X-ray (HXR) flux for the white-light flare. Based on the GOES 1–8 Å SXR flux, the flare lasted approximately 11 minutes, with its start, peak, and end times at 05:54, 06:01, and 06:05 UT, respectively. The impulsive phase, defined by the appearance of HXR emission above 20 keV, spanned from 06:00:03 to 06:01:40 UT and coincided with the impulsive signal in the SXR temporal derivative. During this phase, the white-light flare occurred at around 06:01 UT, suggesting a strong connection between the white-light emission and the nonthermal electron-beam heating. HXI recorded HXR photons primarily below 50 keV, and the potential weak emissions above 50 keV may be submerged by background signals. We used a thick-target model (thick2 in the OSPEX package) and thermal plasma emission model (f_vth) to fit the spectrum, and the result is shown in Figure 1(e). The uncertainties are the best-fit value $3\sigma$ of the data error generated by OSPEX. Note that this is a typical situation where the signal from this C9.3 class flare is comparable to the background of HXI near 30 keV. It is still reasonable to perform spectral analysis due to the relatively small and stable error ($\sim$2%) of HXI background in this case (Z. Li et al. 2025). The low-energy cutoff was $26.6 \pm 5.6$ keV, while the spectral index of the nonthermal component was $7.41 \pm 1.56$, indicating a relatively small population of high-energy electrons. The resulting nonthermal power is $\sim$1.05(0.56–3.55) $\times 10^{27}$ erg s$^{-1}$ during the fitting time interval, with the uncertainty range derived from the full width at half-maximum of the Monte Carlo–generated probability density function. Based on the observed flare kernel size of $\sim$1.58 $\times 10^{17}$ cm$^2$ (region with >2% intensity enhancement in the white box of Figure 2(b)), the source area yields an energy flux density of $\sim$0.67(0.35–2.25) $\times 10^{10}$ erg cm$^{-2}$ s$^{-1}$, which is a relatively low energy flux density compared with previous studies of $\sim 10^{10-12}$ erg cm$^{-2}$ s$^{-1}$ for producing strong white-light emission (M. D. Ding et al. 2003; L. Fletcher et al. 2007; S. Krucker et al. 2011). We also tested double thermal model to fit the spectrum to eliminate the uncertainty raised by spectral fitting approach (S. Krucker & R. P. Lin 2008), but the result shows that the temperature of one of the thermal component needs to be higher than 50 MK, which is almost impossible to appear in a C-class flare (A. Caspi et al. 2014). It consolidates the existence of nonthermal electrons with energies above but near 27 keV due to the steep spectrum. Accordingly, this sub-M-class white-light flare, characterized by a relatively soft nonthermal electron distribution, motivates further investigation into whether factors other than the total released energy contribute to the production of white-light emission.

### 3.2. White-light Emission and Its Time Evolution

Figures 2(a)–(c) present the intensity enhancement maps of the white-light kernels observed in the NVST TiO image, SDO/HMI intensity image, and CHASE continuum image, respectively. The intensity enhancement is calculated as $(I - I_0)/I_0$, where I and $I_0$ are the intensity during (at $\sim$06:01 UT) and immediately before the flare (at $\sim$06:00 UT) at each pixel. The intrinsic variation of the contrast of NVST (1.8%), CHASE (2.4%), and HMI continuum (0.9%) is calculated by measuring the standard deviation of the intensity difference outside of the flare region, as indicated by the white box in Figure 1(c). To emphasize the flaring regions, areas with enhancement below the intrinsic variation are omitted. Clear intensity enhancements were observed in all three bands. The white-light flare signal is particularly prominent in the enhancement maps of the HMI intensity image and the CHASE continuum image, with intensity increases exceeding 8% and 20% in Figures 2(b) and (c), respectively. The white-light emission appears as a brightening region near the PIL of the magnetic $\delta$-sunspot. However, the NVST TiO enhancement map reveals that this brightening region comprises two distinct white-light kernels, labeled K1 and K2, as shown in Figure 2(a). Notably, some filamentary brightenings are also visible, connecting these two kernels, as highlighted by the green dashed lines in Figure 2(a). Based on the NVST TiO image and HMI magnetogram (Figures 2(d) and (e)), the two white-light kernels, K1 and K2, are located at the roots of the penumbra filaments, near the $\delta$-sunspot umbra with positive and negative polarities, respectively. The filamentary brightenings are well aligned with the penumbral structures, as shown by the green dashed lines in Figure 2(d).

To ensure that the observed filamentary brightenings aligned with penumbral structures are not merely due to natural fluctuations in the background photosphere, we selected three representative regions in the TiO images for further analysis, K1 and K2, corresponding to the two white-light flare kernels, and FB, corresponding to the brightened filamentary structure. The intensity evolution of these regions is presented in Figure 2(h). The K1 region shows the strongest enhancement, with an intensity increase exceeding 10%, which is significantly larger than the estimated standard deviation of quiet-Sun TiO fluctuations ($\sim$1.8%). Both the K2 and FB regions show intensity increases of $\sim$5%, also clearly above the background fluctuations. Furthermore, the timing of these intensity enhancements closely follows the flare peak at around 06:01 UT, making it unlikely that they are caused by natural photospheric variability. Therefore, these observations suggest that the brightening observed in the FB region is a real and flare-associated phenomenon, and provide evidence for a photospheric contribution to the white-light emission.

It was noticed that the TiO kernels closely coincided with the HMI continuum kernels, as compared to the green and blue contours in Figure 2(e). This alignment confirms that the TiO intensity enhancement is not an artifact of data noise. Additionally, we noted that the brightening in the CHASE continuum was slightly offset to the east compared to the HMI continuum and NVST TiO images. Given the fact that the white-light flare occurred near the eastern limb of the solar disk, this displacement may be attributed to the higher formation height of the CHASE Fe I continuum relative to the other two, since the Fe I line emission may extend to higher chromosphere due to the flare heating (S. Krucker et al. 2015; J. Hong et al. 2018, 2022). Alternatively, this discrepancy might also result from registration errors, given the substantial differences in spatial resolution among the three instruments. The X-ray image of ASO-S/HXI was reconstructed using the CLEAN algorithm for the energy range of 14–20 keV during 06:00:46 to 06:01:20 UT. The result is shown in Figure 2(f). A primary HXR source was found to align well with the white-light flare region, indicating that the white-light emission was associated with nonthermal electron-beam heating. However, the primary HXR source, with a spatial resolution of 6″.5, lacked sufficient detail to resolve features corresponding to the two white-light kernels. Only two weak HXR sources were identified in extended regions, which is unrelated to the white-light emission.

Figure 2(g) presents the temporal evolution of the integrated intensity in the NVST TiO, SDO/HMI continuum, and CHASE continuum over the white-light flare region. The white-light emission exhibited a noticeable increase around 06:01 UT, coinciding with the peak of the SXR derivative. This timing aligns closely with the peak of HXR emission in the 20–50 keV range, demonstrating the Neupert effect (W. M. Neupert 1968; D. Li et al. 2024). These characteristics suggest that the flare is an impulsive-type white-light flare likely driven by the nonthermal electron beam heating. The normalized white-light emission signals were also found to decrease in intensity in the sequence of CHASE continuum, HMI continuum, and the NVST TiO continuum. The observed intensity differences likely reflect variations in their respective formation heights, temperature sensitivities, and instrumental response functions. The formation height of the Fe I line is estimated to be in the mid-photosphere (200–300 km) for the quiet Sun (A. A. Norton et al. 2006; J. Hong et al. 2022), and it has a more significant response to the flare heating due to the partial contribution from the chromospheric heating (J. Hong et al. 2018). From this view, the CHASE Fe I continuum seems to be the most sensitive to flare heating. Whereas the TiO band is thought to be formed at a deeper height of ∼150 km, and the white-light emission in the TiO band is generally weaker than that of the Fe I line (V. Yurchyshyn et al. 2017).

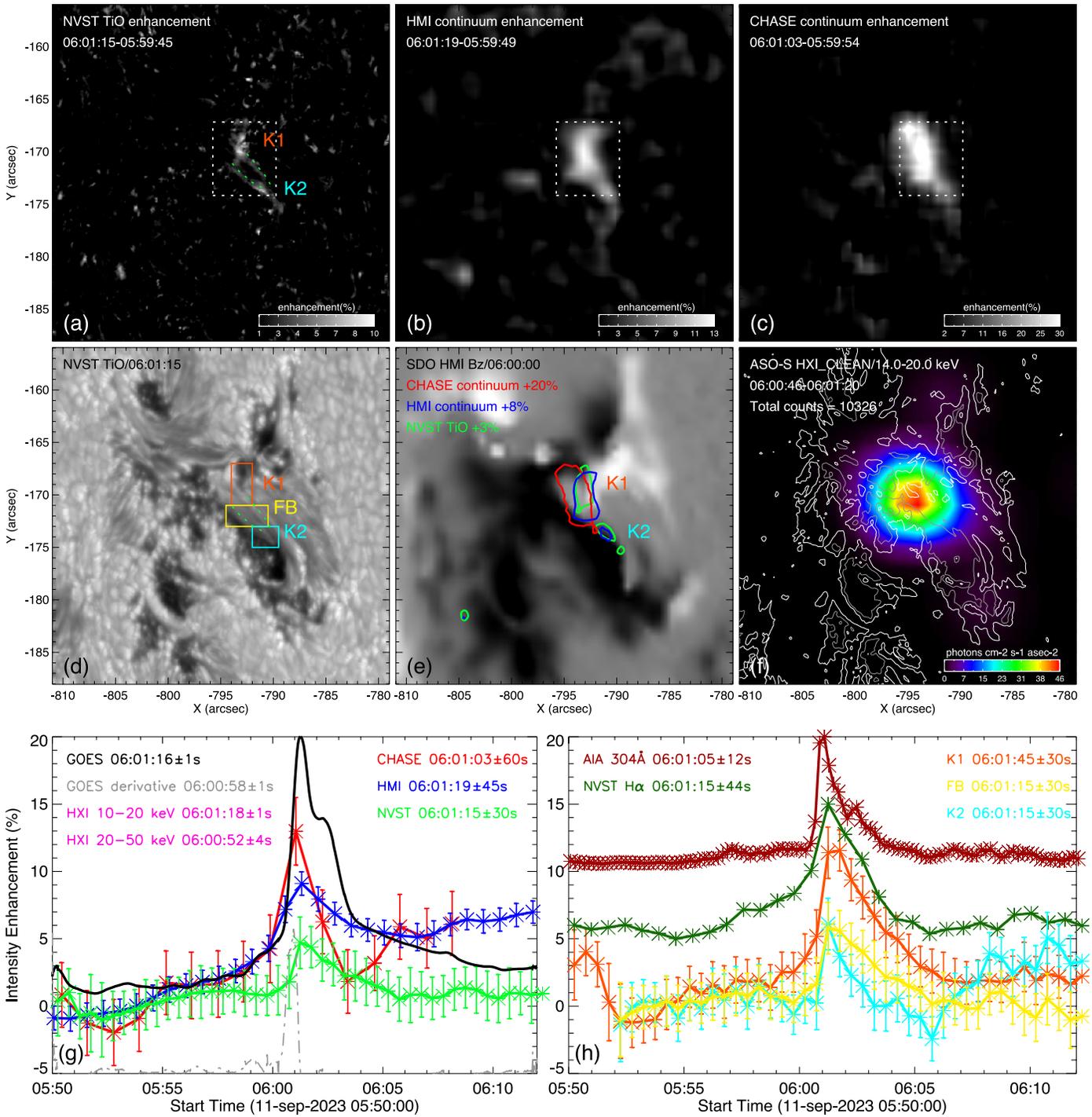

**Figure 2.** Panels (a)–(c): intensity enhancement maps of the NVST TiO, HMI continuum, and CHASE continuum emissions during the white-light flare. Panels (d) and (e): NVST TiO image and HMI magnetogram illustrating the fine structures of the sunspots and their associated magnetic fields in the flaring region. The three colored contours in panel (e) correspond to enhancement levels of 3%, 8%, and 20% in the respective intensity enhancement maps. Panel (f): ASO-S HXI CLEAN-reconstructed map showing X-ray emission in the 14–20 keV range between 06:00:46 and 06:01:20 UT, coinciding with the occurrence of white-light emission. The white and gray contours outline the umbra and penumbra boundaries of sunspots in the NVST TiO image. Panel (g): light curves of the NVST TiO, HMI continuum, and CHASE continuum emissions integrated over the white dashed box region in panels (a)–(c), respectively. The GOES soft X-ray flux and its time derivative are overplotted in arbitrary units for comparison. Panel (h): light curves of the K1, K2, and FB regions integrated over the orange, yellow, and cyan boxes in NVST TiO images, respectively. The light curves of the AIA 304 Å and NVST Hα integrated over the white dashed box region in Figures 3(h) and (k) are overplotted in arbitrary units for comparison. The labeled times denote the peak times of each curve. The error bars for each curve are the standard deviation of the intensity change taken from the nonflare region (white box in Figure 1(c)).

To investigate the possible temporal relationship between emissions at different atmospheric heights, we further analyzed the intensity evolution of the flare kernels in AIA 304 Å, NVST Hα, and NVST TiO bands. As shown in Figure 2(h), the overall profiles suggest a offset in time that the AIA 304 Å reached its peak slightly earlier than the NVST Hα, which is then followed by the photospheric TiO enhancement. Specifically, the intensity peak in the AIA 304 Å likely occurred before 06:01:05 UT, as the rate of increase slows significantly by that time. The Hα intensity reached its peak around 06:01:15 UT, while the TiO intensity peaks later, after 06:01:15 UT. Notably, in the brightest white-light kernel region K1, the TiO peak occurs around 06:01:45 UT. Moreover, the AIA 304 Å peak closely aligns with the peak of HXI with energy range (20–50 keV), while the Hα and TiO peaks are more consistent with the peak of HXI with energy range (10–20 keV). This further supports the idea that different atmospheric layers respond at different times and possibly through distinct mechanisms of energy deposition. By comparing the peak times of the light curves, a possible time delay of approximately 10–50 s between 304 Å and TiO emissions is suggested. However, this delay remains uncertain, as it cannot be well resolved given the 30 s cadence of the NVST TiO data. Interestingly, a recent study by J. Lörinčík et al. (2025) using very high-cadence IRIS observations reported a time delay between flare peak in the transition region and chromosphere, and suggested that Alfvén waves is one possible energy transport mechanism that could account for the observed delay. In this event, if assuming a chromospheric thickness of ∼2000 km and an Alfvén speed of ∼50 km s$^{-1}$, the resulting travel time ∼40 s is consistent with the suggested delay. However, this conclusion is not definitive, and high-cadence data are needed to more accurately resolve time delays and investigate energy transport mechanisms in flares.

### 3.3. Flare Eruption and Spectral Observations

To figure out how this flare was initiated and evolved, we further investigated the photospheric and chromospheric dynamics before and during the flare. Figures 3(a) and (b) compare the active region approximately 50 minutes before and just prior to its onset. It was clear that a pair of sunspots, labeled P and N, exhibited pronounced shear motion before the flare. The shearing velocity reached up to 0.5 km s$^{-1}$, which likely facilitated the continuous transport of magnetic energy into the upper atmosphere, eventually driving the flare eruption (L. Li & J. Zhang 2009; Y. Zhang et al. 2022). As shown in the magnetograms in Figures 3(c) and (d), the northward movement of the positive-polarity spot P brought it into contact with a region of negative polarity ahead. This interaction led to persistent magnetic cancellation in the region, highlighted by the green dashed circle in Figure 3(d). The magnetic cancellation, which began tens of minutes before the flare, was likely a critical factor in triggering the eruption.

The chromospheric evolution is depicted through sequences of NVST Hα and AIA 304 Å images in Figure 3. By comparing the Hα images at 05:10 and 05:55 UT (Figures 3(e) and (f)), a filament can be seen forming directly above the region of photospheric shear motion. This filament became the primary structure for the subsequent flare eruption. During its formation, a small cusp-shaped brightening appeared below the filament, spatially aligned with the region of magnetic cancellation. This alignment suggests that magnetic reconnection occurred at this location. At 06:00 UT (Figure 3(g)), the southern part of the filament thread became activated just before the flare, accompanied by significant brightening at the footpoints of the filament. By 06:01 UT (Figure 3(h)), the filament thread began to erupt, triggering the C9.3 white-light flare. Comparing the contours of the Hα emission (orange) and TiO enhancement (cyan) reveals that the northern Hα kernel and white-light kernel overlapped closely, whereas the southern kernels did not align perfectly. Additionally, the distance between the white-light kernels was shorter than that between the Hα kernels, implying that the magnetic system associated with the white-light kernels was located at a lower height than the system involved in the Hα kernels. The filament eruption also generated a solar jet along the overlying loops, as illustrated in the AIA 304 Å images in Figures 3(i)–(l). At 06:01 UT (Figure 3(k)), two brightening points appeared at the base of the solar jet. These brightening points coincided both temporally and spatially with the white-light kernels, further linking the flare dynamics to the observed white-light emissions.

CHASE provides spectral observations of the flare eruption in both the photosphere and chromosphere by capturing the Fe I line and Hα line, respectively. Figure 4(b) compares the Fe I line profiles in the white-light flare region 5 minutes before, during, and 5 minutes after the flare peak. The Fe I line was significantly influenced by the flare, with its line center intensity increasing by approximately 46% during the event. The continuum of the Fe I line exhibited enhancement and a notable blue asymmetry, likely linked to dynamic processes indicated by the pronounced red asymmetry in the nearby Hα line (Figure 4(d)). The enhanced chromospheric activities, such as chromospheric condensation or plasma downflows, could contribute to the observed blue asymmetry in the Fe I line, suggesting that chromospheric processes rather than photospheric motions are primarily responsible for the Fe I blue-wing asymmetry. Throughout the flare, the Fe I line maintained an absorption profile. To analyze this further, the Fe I line profiles were fitted using a Gaussian term combined with a linear term. The result reveals line broadening during the flare, with the line width increasing from 1.22 $W_{qs}$ (quiet-Sun line width) to 1.32 $W_{qs}$ and subsequently returning to 1.20 $W_{qs}$ after the flare. The Doppler velocity of the Fe I line exhibited a weak blueshift of $-1$ km s$^{-1}$ during the flare, but this shift appeared unrelated to the flare since similar blueshifts were present before and after the event. In contrast, the Hα line in the white-light flare region transitioned from absorption to enhanced emission during the flare, as shown in Figure 4(d).

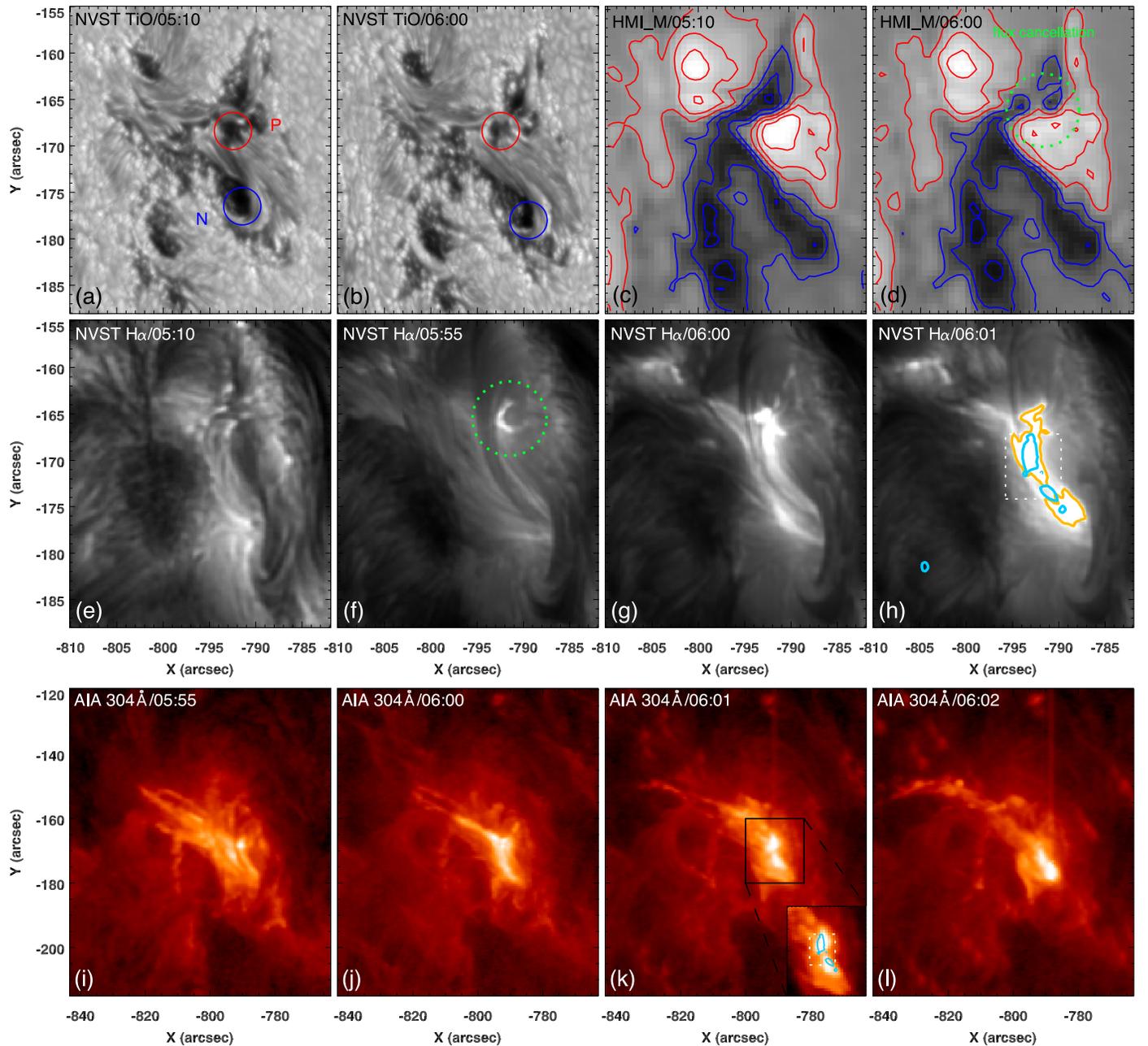

**Figure 3.** Panels (a)–(d): NVST TiO images and HMI LOS magnetograms depicting the photospheric evolution leading up to the flare. The red and blue circles mark the spots P and N, which exhibited significant shearing motion. The contour levels in panels (c) and (d) represent the magnetic field strengths of ±200, 500, and 800 G. Panels (e)–(h): NVST Hα images showing the chromospheric evolution toward the flare. The green dotted circle in panels (d) and (f) highlights the magnetic cancellation region. The orange and cyan contours in panel (h) outline the flare kernels observed in the chromosphere and photosphere, respectively. Panel (i)–(l): AIA 304 Å images with a larger field of view, showing the solar jet associated with the flare.

The emission profile exhibited a centrally reversed shape with an evident red asymmetry. To quantify these changes, Gaussian fits were also applied to the Hα contrast line profiles (solid colored curves), which were derived by subtracting the reference Hα profile from the quiet Sun (black dotted line) from the Hα profile in the white-light flare region (colored cross symbols). The fitting results indicate that the Hα line also broadened significantly during the flare, with its line width increasing from 0.62 $W_{qs}$ (line width in the quiet Sun) to 1.46 $W_{qs}$ during the flare and then decreasing to 0.73 $W_{qs}$ afterward. The Doppler velocity of the Hα line shows a redshift of 5.6 km s$^{-1}$ during the flare, compared to relatively small shifts before (0.8 km s$^{-1}$) and after (−0.9 km s$^{-1}$) the flare. As suggested by many previous studies (M. D. Ding & C. Fang 1996; D. Li et al. 2023; D.-C. Song et al. 2023), the observed red asymmetry and Doppler redshift in the Hα line are likely due to the formation of its red wing in a chromospheric condensation region characterized by downflows.

### 3.4. Sudden Photospheric Vortex and Magnetic Field Amplification

Interestingly, a sudden photospheric vortex emerged at the northern white-light flare kernel immediately after the flare peak. Figures 5(a)–(d) show the sequence of NVST TiO images, depicting the photospheric evolution in the white-light flare region. Two tiny spots appeared within the northern white-light flare kernel, as indicated by the blue arrows in Figure 5(b). These spots then underwent a counterclockwise rotational motion around each other before eventually merging into a single pore, highlighted by the red arrow in Figure 5(d). This dynamic process is more vividly illustrated in the accompanying animation in Figure 6 in the Appendix. The appearance of the spots and their vortex motion is strongly associated with the white-light flare, as they emerged during the flare and persisted for only about 10 minutes following the flare peak.

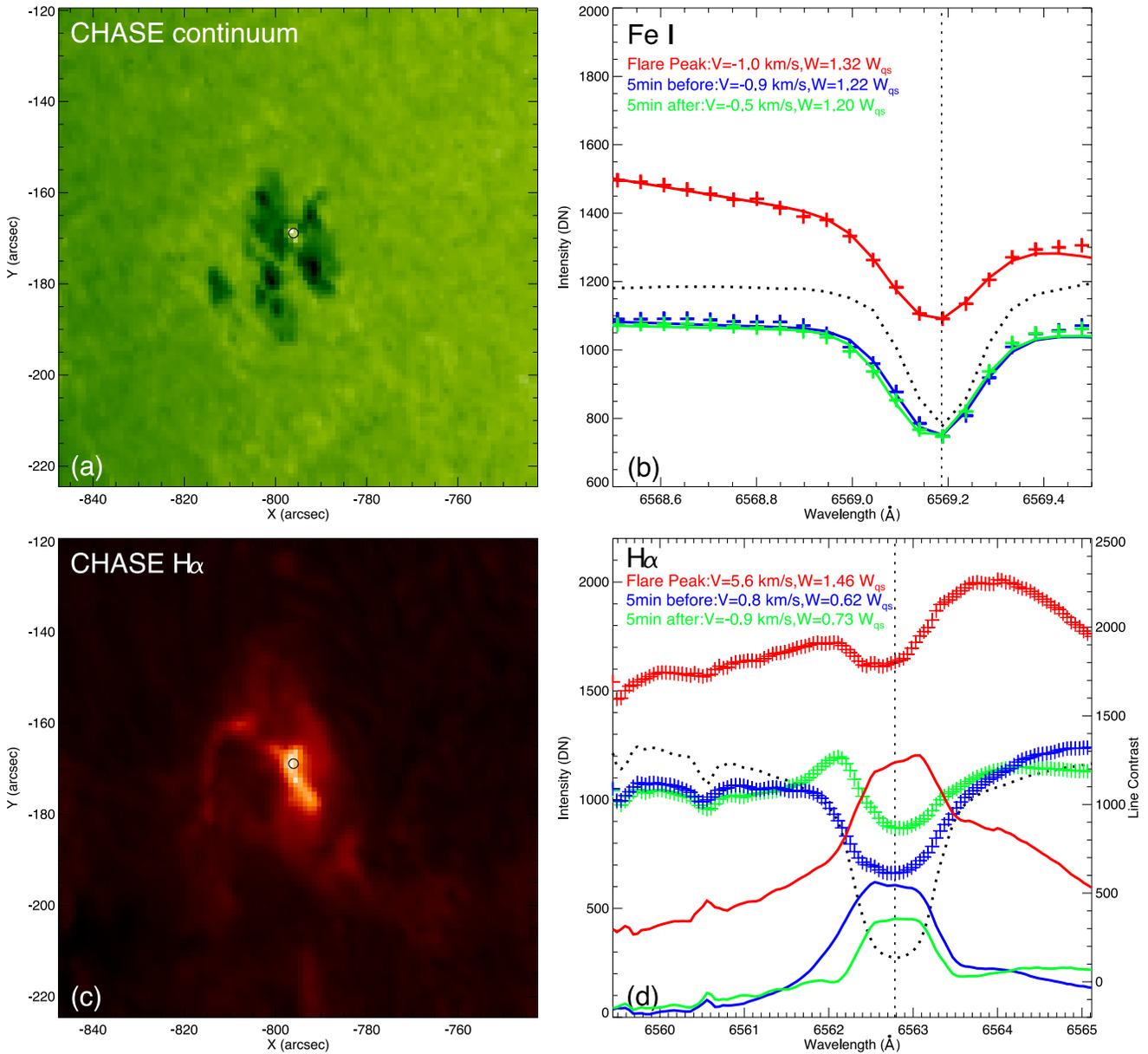

**Figure 4.** Panels (a) and (b): CHASE continuum image and the Fe I line profiles in the white-light flare region (marked by the black circle). The red, blue, and green cross symbols represent the Fe I line profiles during, before, and after the flare. The corresponding solid curves are the Gaussian fits of the line profiles combined with a linear term. For reference, the black dotted curve shows the Fe I line profile in a nonflaring region. Panels (c) and (d): CHASE H$\alpha$ image and its line profiles (colored cross symbol) in the white-light flare region. The colored solid curves represent the contrast profiles relative to the reference H$\alpha$ line profile (black dotted line) from a nonflaring region. The vertical dashed lines in panels (b) and (d) indicate the line centers of the Fe I and H$\alpha$ lines, respectively.

The photospheric flow fields immediately before and after the flare were obtained by applying the dense optical flow method to the NVST TiO images, with the results shown in Figures 5(e) and (f). Before the flare, the flow fields at the northern white-light flare kernel were well aligned along the penumbra fibrils, displaying a characteristic penumbral flow pattern (Figure 5(e)). In contrast, after the flare, this penumbral flow transformed into a counterclockwise vortex flow, with the local penumbral fibrils evolving into a pore (Figure 5(f)). Notably, the vortex flow was centered precisely at the location of the northern white-light flare kernel. Meanwhile, the flare-related velocity perturbations are also present in the southern flare kernel and in the filamentary-brightening region between the two kernels, as evident in the comparison between Figures 5(e) and (f), and further supported by the animation provided in the Appendix. In these two regions, however, the velocity variations are significantly weaker than those in the northern flare kernel.

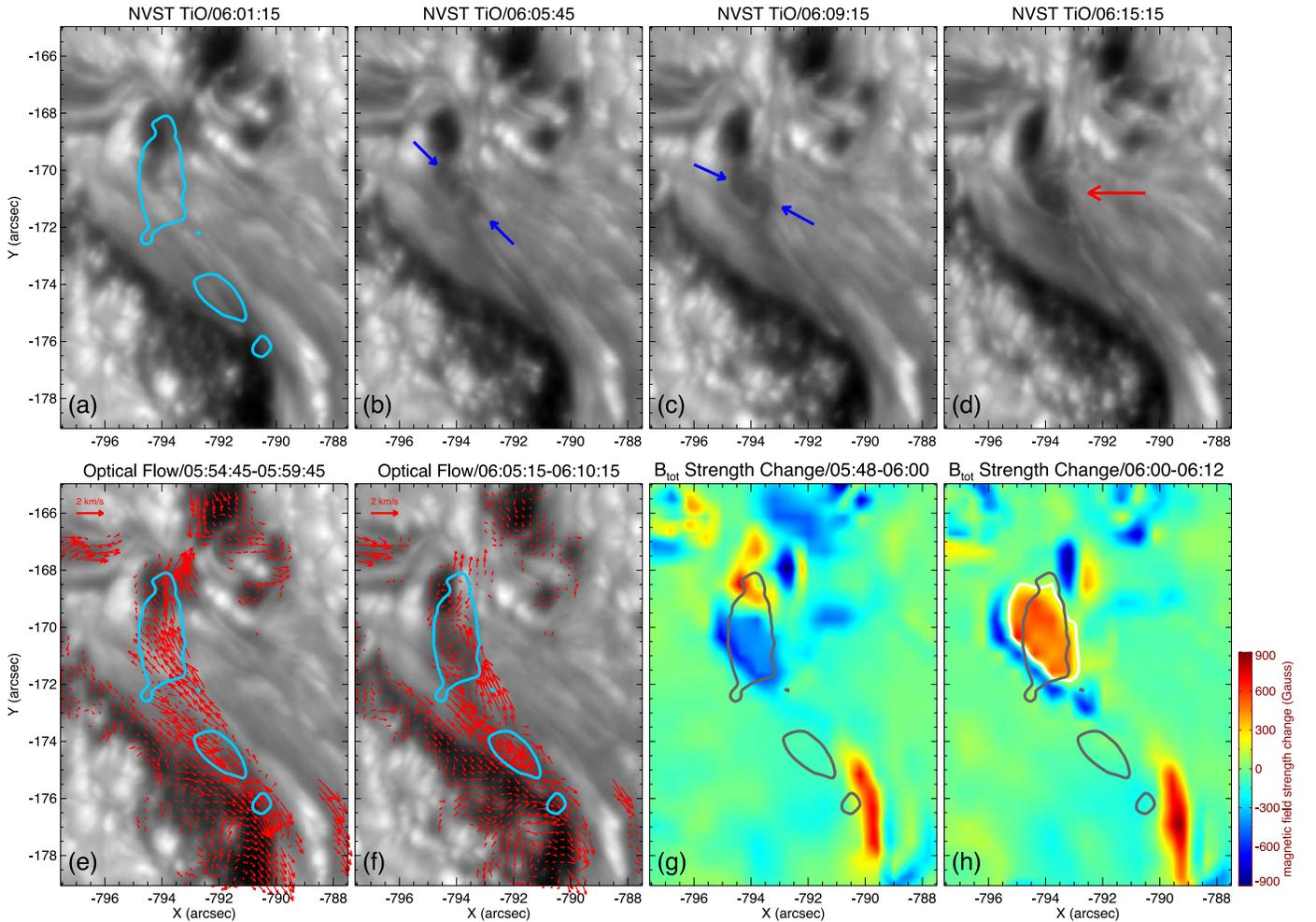

**Figure 5.** Panels (a)–(d): NVST TiO images illustrating the sudden formation of a small pore in the white-light flare region. Panels (e) and (f): photospheric flow fields derived using the optical flow method applied to NVST TiO images, showing the velocity patterns 5 minutes before and after the white-light flare. Panels (g) and (h): maps of the total magnetic field strength before and after the white-light flare, highlighting magnetic field changes in the flare region. The blue and gray contours overlaid in panels (a) and (e)–(h) denote the locations of the white-light flare kernels. The white contour in panel (h) marks the magnetic field amplification region with enhanced strength greater than 150 G.

Numerous studies have shown that sudden structural changes in sunspots during a flare are often attributed to abrupt variations in the photospheric magnetic fields (see H. Wang & C. Liu 2015 for a review). To further investigate this, the changes in the magnetic field strength ($B_{tot}$) immediately before and after the flare were analyzed, as presented in Figures 5(g) and (h). Around the northern white-light flare kernel, the average magnetic field strength, which had been gradually decreasing prior to the flare, experienced a sudden increase after the flare. We estimate the magnetic field amplification around the northern white-light flare kernel by selecting the region where the field enhancement is greater than 150 G. As shown by the white contour in Figure 5(h), most of the magnetic amplification region overlaps with the north white-light flare kernel. The result shows that its average magnetic field strength increased from ∼650 to ∼1050 G. Such a process is clearly associated with the flare, which suggests that the energy released during the flare might be transferred to the photosphere and stored as magnetic energy.

However, no significant changes in magnetic field strength were observed at the southern white-light flare kernel. Instead, an elongated region of enhanced magnetic fields was evident adjacent to the southern white-light flare kernel both before and after the flare. It is likely a localized process unrelated to the flare, driven by the shearing motion of the southern umbra, which can continuously amplify the magnetic fields in the region ahead of it (Z. Xu et al. 2020, 2022; C. Jin et al. 2024).

## 4. Conclusions and Discussion

In this study, we presented a detailed analysis of a C9.3 white-light flare that occurred in AR 13431 on 2023 September 11. The white-light emission was concentrated in the central penumbra of a δ-type sunspot and was effectively captured by the NVST TiO, HMI continuum, and CHASE Fe I continuum images. The emission exhibited two brightening kernels located at opposite magnetic polarities, connected by filamentary brightenings. The localized brightening of these sunspot fine structures suggests a photospheric contribution to the white-light emission. Spectral observations of the CHASE Fe I line, which showed significant intensity enhancement and line broadening during the flare, further corroborate this conclusion. The timing of the white-light emission peak closely aligned with the peak of the HXR emission, indicating a strong association with the nonthermal electron beam heating process. However, ASO-S HXI data revealed no detection of electrons with energies exceeding 50 keV in this C9.3 flare. This suggests that additional mechanisms might be required to facilitate electron penetration into the deeper layers of the solar atmosphere.

Alfvén wave dissipation has been proposed as a mechanism to assist in the deposition of energetic electrons into deeper layers of the solar atmosphere. In this event, the impact of the flare on the solar photosphere was distinctly observed, manifesting as sudden vortex flows and the associated magnetic field amplification around the region of the white-light flare kernel. Since the vortex flows consisted of two small spots with the same polarity rotating around each other, they resemble a bundle of magnetic flux tubes twisting during the flare. Hence, we infer that the sudden vortex flows represent evidence of a flare-induced Alfvén wave pulse propagating the twist from the upper atmosphere to the photosphere, while the observed magnetic field amplification results from the deposition of Alfvén wave pulse energy. To estimate the magnetic energy density required to enhance the local magnetic field, the equation $\delta e_B = \frac{B_t^2 - B_0^2}{2\mu_0}$ was used, where $\mu_0$ is the magnetic permeability in vacuum, $B_0 = 650$ G is the average magnetic field strength around the white-light kernel (white contour in Figure 5(h)) before the flare, and $B_t = 1050$ G is the strength after the flare. Assuming a photospheric height of 500 km, the volume of the magnetic field amplification region is about $1.52 \times 10^{24}$ cm$^3$. Using these values, the required magnetic energy was estimated to be about $4.11 \times 10^{28}$ erg. This amount of energy is comparable to previous statistical findings on the transfer of magnetic energy across the photosphere during relatively low-class flares (Y. Bi et al. 2018). We also note that the southern flare kernel does not exhibit significant magnetic field enhancement, and the associated velocity perturbations are markedly weaker compared to those observed in the northern kernel. We attribute this to the highly asymmetric structure of the filament eruption in this event, as illustrated in Figure 3(j). The eruption source is located above the northern flare kernel, lying closer than to the southern kernel. As a result, the Alfvén wave pulse exerts a stronger impact on the northern region. This interpretation is consistent with the observation in Figure 2(h), where the white-light emission from the northern kernel K1 is significantly more intense than that from the southern kernel K2.

The nonthermal energy flux derived from ASO-S/HXI spectral fitting during 06:00:45 to 06:01:33 UT is $\sim 1.05 \times 10^{27}$ erg s$^{-1}$, the thermal energy of the plasma with an assumed volume of $10^{27}$ cm$^3$ is $2.5 \times 10^{29}$ erg. The nonthermal energy is therefore a small portion ($\sim$one-sixth) of the total released energy. Furthermore, the deposited magnetic energy is of the same order of magnitude as the nonthermal electron energy, suggesting a potential relationship between Alfvén waves and the electron acceleration process. As in many previous studies (L. Fletcher & H. S. Hudson 2008; M. Mathioudakis et al. 2013), the magnitude of Poynting flux carried by Alfvén waves is estimated by combining the Alfvén speed in the corona with the magnetic field perturbations observed at the photosphere. Here, we clearly observed the velocity perturbations associated with Alfvén wave pulses in the photosphere, with an average amplitude of $\sim 0.6$ km s$^{-1}$ in regions of magnetic field enhancement (white contour in Figure 5(h)). This corresponds to a magnetic field perturbation of $\sim 100$ G with a typical photospheric mass density of $\rho = 2 \times 10^{-4}$ kg m$^{-3}$, as estimated using the relation $\delta B_\perp = \sqrt{\mu_0 \rho} \cdot v_\perp$, which is derived from the linearized MHD equations for an incompressible Alfvén wave propagating along a uniform background magnetic field. Assuming the Alfvén wave pulse originates in the corona, where the typical Alfvén wave speed is $\sim 1000$ km s$^{-1}$, the Poynting flux was estimated as $S \approx v_A \cdot \frac{\delta B_\perp^2}{4\pi} \approx 8 \times 10^{10}$ erg cm$^{-2}$ s$^{-1}$. Over a flaring area of $\sim 1.58 \times 10^{17}$ cm$^2$ (region with >2% intensity enhancement in the white box of Figure 2(b)) and a duration of 5 minutes, the total energy carried by the Alfvén wave pulse is $\sim 3.79 \times 10^{30}$ erg, which is more than sufficient to account for the observed photospheric magnetic energy deposition ($\sim 4.11 \times 10^{28}$ erg). The energy flux of the Alfvén wave pulse is $\sim 1.26 \times 10^{28}$ erg s$^{-1}$, which also well exceeds the nonthermal energy flux of $\sim 1.05(0.56$–$3.55) \times 10^{27}$ erg s$^{-1}$ during the flare peak.

Based on the energy estimation, only a small fraction of the Alfvén wave energy appears to be transported to the photosphere. It is consistent with previous studies showing that Alfvén waves can be efficiently dissipated in the chromosphere, contributing to localized atmospheric heating (G. S. Kerr et al. 2016; J. W. Reep & A. J. B. Russell 2016; J. W. Reep et al. 2018). We also noted that the dissipation of Alfvén waves is frequency-dependent. The higher-frequency Alfvén waves tend to dissipate more efficiently in the chromosphere, while lower-frequency Alfvén waves are more likely to propagate through to a deeper layer. In this event, the Alfvén wave pulse observed in the photosphere exhibits a relatively long period on the order of several minutes, supporting the idea that lower-frequency Alfvén waves might be able to travel to the photosphere.

However, direct evidence of Alfvén wave dissipation leading to atmospheric heating was not obtained in this event. Our observations only suggest that the Alfvén wave pulse may transfer energy down to the photosphere, where some of the energy is converted into localized magnetic energy. If photospheric heating indeed occurs, an efficient energy transport mechanism would be required to carry flare-released energy down to the photosphere. According to the scenario proposed by L. Fletcher & H. S. Hudson (2008), the Alfvén wave pulse can rapidly transfer energy and magnetic field variations from the corona to the lower atmosphere. They further suggested that these pulses could generate field-aligned electric fields capable of accelerating electrons to heat the deeper layers of the Sun. In this event, the rapid magnetic field changes were clearly observed, but the associated electron acceleration process could not be confirmed with the current observations. Nevertheless, given the close temporal and spatial correlation between the Alfvén wave pulse and the white-light emission, it remains plausible that the Alfvén wave pulse contributed to nonthermal electron beam heating in the deeper photosphere.

If we assume that the white-light emission indeed has a contribution from the photosphere, the radiative back-warming mechanism is also a plausible mechanism, acting as a secondary effect of nonthermal electron beam heating. In this event, the white-light flare kernels were spatially coincident with chromospheric brightening features, particularly those observed in the AIA 304 Å band. However, the white-light flare kernels are not observed to be broader than the chromospheric brightenings, considering that the chromospheric brightening emits downward isotropically as a radiation source. This may be explained by the fact that a circular chromospheric radiation source produces a Gaussian-like distribution of heating power in the underlying photosphere, with the maximum occurring directly below the source center. Only the central area receives sufficient energy to raise the photospheric temperature to levels detectable in the white-light continuum. The time delay between the chromospheric brightening and the photospheric white-light emission was found to be within 1 minute, although this measurement was constrained by the time resolution of the photospheric data.

The white-light emission during flares may also be partially attributed to chromospheric heating. R. Joshi et al. (2021) investigated a mini-flare event and reported Balmer continuum enhancements observed by IRIS, concluding that the continuum emission was influenced by chromospheric heating resulting from nonthermal electron energy deposition. Their study shares similarities with our investigation, as both events involve low-class flares associated with white-light emission. In our case, a chromospheric contribution to the white-light emission is also plausible, given the strong spatial correspondence between the white-light flare kernels and chromospheric brightenings in the AIA 304 Å image. However, the high-resolution TiO observations revealed the white-light enhancement features resembling penumbral fibrils, suggesting that a photospheric contribution is also present. Accordingly, our study supports the interpretation that both the photosphere and chromosphere may contribute simultaneously to the observed white-light emission (G. S. Kerr & L. Fletcher 2014; L. Kleint et al. 2016; J. Hong et al. 2018).

This study demonstrates that nonthermal electrons with energies below 50 keV can contribute to heating the solar photosphere during a C-class flare. Additionally, compelling evidence indicates that a large-scale Alfvén wave pulse transported energy to the photosphere, suggesting the potential role of an Alfvén wave dissipation mechanism. However, the acceleration of nonthermal electrons by the Alfvén wave pulse and the damping of the Alfvén wave cannot be confirmed based on the current observations. Further numerical simulations are required to validate the hypothesis that Alfvén wave dissipation can facilitate low-energy nonthermal electrons in heating the deeper layers of the Sun.

### Acknowledgments

The authors would like to thank the referee for a very careful reading and suggestions that improved the quality of the manuscript. We are also grateful for the insightful discussions with Dr. Dechao Song. We thank the NVST, SDO, ASO-S, and CHASE teams for providing the data. This work is supported by the Strategic Priority Research Program of the Chinese Academy of Science, grant No. XDB0560000, the National Science Foundation of China (NSFC) under grants 12203097, 12325303, 12433010, 12173084, 12273108, 12333010, and 12273106, the National Key R&D Program of China 2022YFF0503002, the Yunnan Key Laboratory of Solar Physics and Space Science under the number 202205AG070009, the Yunnan Science Foundation of China (202301AT070349 and 202301AT070347), the Prominent Postdoctoral Project of Jiangsu Province (2023ZB304), and the Young Elite Scientists Sponsorship Program by Yunnan Association for Science and Technology.

### Appendix

An online animation (Figure 6) is available, showing the photospheric evolution of AR 13431 from 05:10 to 06:35 UT on 2023 September 11, as observed in the NVST TiO images.

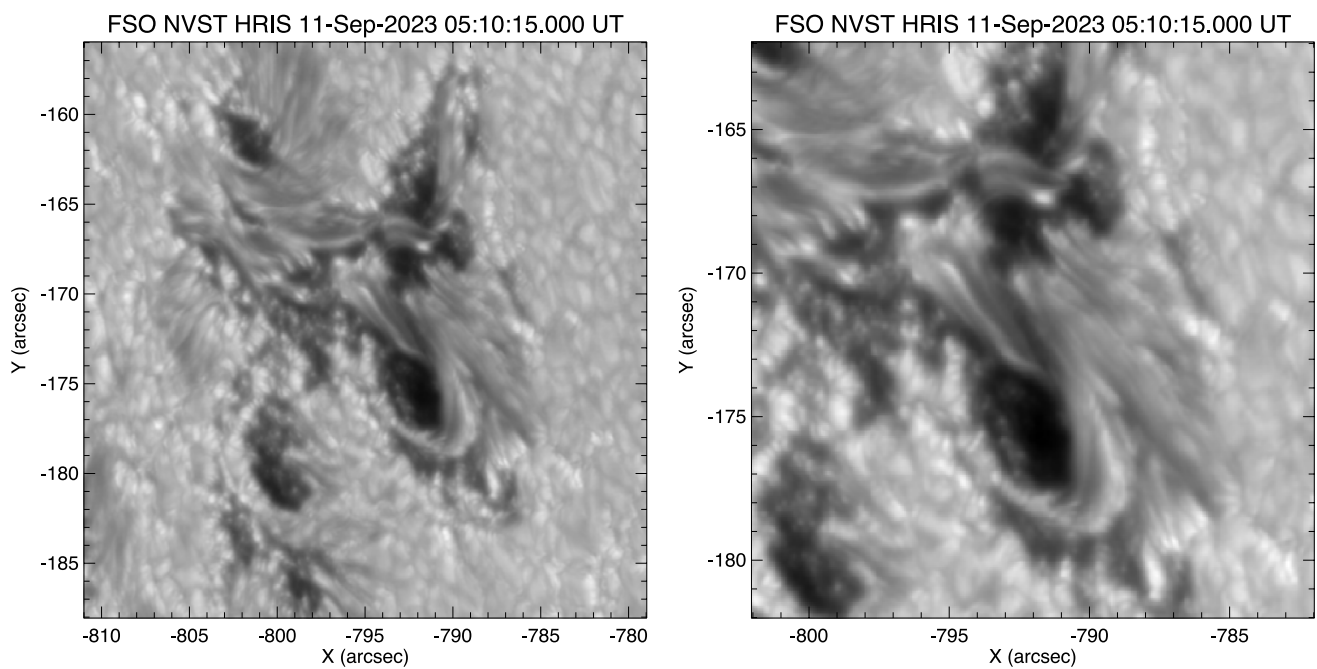

**Figure 6.** An animation showing the photospheric evolution of AR 13431 from 05:10 to 06:35 UT on 2023 September 11, as observed in the NVST TiO images. The left panel presents a large FOV, displaying the white-light flare event continuously with a cadence of 30 s. The right panel focuses on the white-light flare region in a closer FOV, where each frame is a lucky image, selected to correspond closely in time to the frames shown in the left panel.
(An animation of this figure is available in the online article.)

## ORCID iDs

Zhe Xu ⓘ https://orcid.org/0000-0002-9121-9686
Xiaoli Yan ⓘ https://orcid.org/0000-0003-2891-6267
Zhentong Li ⓘ https://orcid.org/0000-0002-4230-2520
Liheng Yang ⓘ https://orcid.org/0000-0003-0236-2243
Zhike Xue ⓘ https://orcid.org/0000-0002-6526-5363
Jincheng Wang ⓘ https://orcid.org/0000-0003-4393-9731

## References

Aboudarham, J., & Henoux, J. C. 1987, A&A, 174, 270
Bi, Y., Liu, Y. D., Liu, Y., et al. 2018, ApJ, 865, 139
Brown, J. C., Turkmani, R., Kontar, E. P., et al. 2009, A&A, 508, 993
Castellanos Durán, J. S., & Kleint, L. 2020, ApJ, 904, 96
Cai, Y., Hou, Y., Li, T., et al. 2024a, ApJ, 975, 69
Cai, Y., Xiang, Y., & Ji, K. 2024b, ApJ, 977, 186
Caspi, A., Krucker, S., & Lin, R. P. 2014, ApJ, 781, 43
Cliver, E. W., Petrie, G. J. D., & Ling, A. G. 2012, ApJ, 756, 144
Chen, H., Fletcher, L., Zhou, G., et al. 2024, ApJ, 976, 207
Chen, P.-F., Fang, C., & Ding, M.-D. D. 2001, ChJAA, 1, 176
Ding, M. D., & Fang, C. 1996, SoPh, 166, 437
Ding, M. D., Fang, C., & Yun, H. S. 1999, ApJ, 512, 454
Ding, M. D., Liu, Y., Yeh, C.-T., et al. 2003, A&A, 403, 1151
Emslie, A. G., & Sturrock, P. A. 1982, SoPh, 80, 99
Fang, C., Chen, P.-F., Li, Z., et al. 2013, RAA, 13, 1509
Fletcher, L., Hannah, I. G., Hudson, H. S., et al. 2007, ApJ, 656, 1187
Fletcher, L., & Hudson, H. S. 2008, ApJ, 675, 1645
Gan, W. Q., & Mauas, P. J. D. 1994, ApJ, 430, 891
Gan, W. Q., Rieger, E., Zhang, H. Q., et al. 1992, ApJ, 397, 694
Gan, W., Zhu, C., Deng, Y., et al. 2023, SoPh, 298, 68
Gong, L., Yan, X., Liang, H., et al. 2024, MNRAS, 530, 3897
Farnebäck, G. 2003, LNCS, 2749, 363
Hao, Q., Guo, Y., Dai, Y., et al. 2012, A&A, 544, L17
Hao, Q., Yang, K., Cheng, X., et al. 2017, NatCo, 8, 2202
Hong, J., Ding, M. D., Li, Y., et al. 2018, ApJL, 857, L2
Hong, J., Qiu, Y., Hao, Q., et al. 2022, A&A, 668, A9
Hudson, H. S. 1972, SoPh, 24, 414
Hudson, H. S. 2011, SSRv, 158, 5
Hudson, H. S. 2016, SoPh, 291, 1273
Hudson, H. S., Wolfson, C. J., & Metcalf, T. R. 2006, SoPh, 234, 79
Jess, D. B., Mathioudakis, M., Crockett, P. J., et al. 2008, ApJL, 688, L119
Ji, K., Liu, H., Jin, Z., Shang, Z., & Qiang, Z. 2019, ChSBu, 64, 1738
Jing, Z., Li, Y., Feng, L., et al. 2024, SoPh, 299, 11
Jin, C., Zhou, G., Ji, H., et al. 2024, ApJ, 975, 46
Joshi, R., Schmieder, B., Heinzel, P., et al. 2021, A&A, 654, A31
Kerr, G. S., & Fletcher, L. 2014, ApJ, 783, 98
Kerr, G. S., Fletcher, L., Russell, A. J. B., et al. 2016, ApJ, 827, 101
Kleint, L., Heinzel, P., Judge, P., et al. 2016, ApJ, 816, 88
Krucker, S., Hudson, H. S., Jeffrey, N. L. S., et al. 2011, ApJ, 739, 96
Krucker, S., & Lin, R. P. 2008, ApJ, 673, 1181
Krucker, S., Saint-Hilaire, P., Hudson, H. S., et al. 2015, ApJ, 802, 19
Kuhar, M., Krucker, S., Martínez Oliveros, J. C., et al. 2016, ApJ, 816, 6
Lemen, J. R., Title, A. M., Akin, D. J., et al. 2012, SoPh, 275, 17
Li, C., Fang, C., Li, Z., et al. 2022, SCPMA, 65, 289602
Li, D., Dong, H., Chen, W., et al. 2024, SoPh, 299, 57
Li, D., Li, C., Qiu, Y., et al. 2023, ApJ, 954, 7
Li, L., & Zhang, J. 2009, ApJL, 706, L17
Li, Q., Li, Y., Su, Y., et al. 2024a, SoPh, 299, 73
Li, Y., Jing, Z., Song, D.-C., et al. 2024c, ApJL, 963, L3
Li, Y., Liu, X., Jing, Z., et al. 2024b, ApJL, 972, L1
Li, Z., Su, Y., Liu, W., et al. 2025, SoPh, 300, 56
Liu, Z., Xu, J., Gu, B.-Z., et al. 2014, RAA, 14, 705
Lörinčík, J., Polito, V., Kerr, G. S., et al. 2025, arXiv:2504.10619
Machado, M. E., Emslie, A. G., & Avrett, E. H. 1989, SoPh, 124, 303
Martínez Oliveros, J.-C., Hudson, H. S., Hurford, G. J., et al. 2012, ApJL, 753, L26
Mathioudakis, M., Jess, D. B., Erdélyi, R., et al. 2013, SSRv, 175, 1
Metcalf, T. R., Alexander, D., Hudson, H. S., et al. 2003, ApJ, 595, 483
Neidig, D. F. 1989, SoPh, 121, 261
Neupert, W. M. 1968, ApJL, 153, L59
Norton, A. A., Graham, J. P., Ulrich, R. K., et al. 2006, SoPh, 239, 69
Penn, M., Krucker, S., Hudson, H., et al. 2016, ApJL, 819, L30
Pesnell, W. D., Thompson, B. J., & Chamberlin, P. C. 2012, SoPh, 275, 3
Qiu, Y., Rao, S., Li, C., et al. 2022, SCPMA, 65, 289603
Reep, J. W., & Russell, A. J. B. 2016, ApJL, 818, L20
Reep, J. W., Russell, A. J. B., Tarr, L. A., et al. 2018, ApJ, 853, 101
Schou, J., Scherrer, P. H., Bush, R. I., et al. 2012, SoPh, 275, 229
Song, D.-C., Dominique, M., Zimovets, I., et al. 2025, ApJL, 983, L41
Song, D.-C., Tian, J., Li, Y., et al. 2023, ApJL, 952, L6
Song, Y., & Tian, H. 2018, ApJ, 867, 159
Song, Y., Tian, H., Zhu, X., et al. 2020, ApJL, 893, L13
Song, Y. L., Tian, H., Zhang, M., et al. 2018, A&A, 613, A69
Su, Y., Zhang, Z., Chen, W., et al. 2024, SoPh, 299, 10
Su, Y., Zhang, Z., Gan, W., et al. 2022, in Handbook of X-ray and Gamma-ray Astrophysics, ed. C. Bambi & A. Santangelo (Singapore: Springer)
Varady, M., Karlický, M., Moravec, Z., et al. 2014, A&A, 563, A51
Wang, H., & Liu, C. 2010, ApJL, 716, L195
Wang, H., & Liu, C. 2015, RAA, 15, 145
Watanabe, K., Krucker, S., Hudson, H., et al. 2010, ApJ, 715, 651
Wheatland, M. S., Melrose, D. B., & Mastrano, A. 2018, ApJ, 864, 159
Xu, Y., Cao, W., Liu, C., et al. 2006, ApJ, 641, 1210
Xu, Z., Ji, H., Hong, J., et al. 2022, A&A, 660, A55
Xu, Z., Ji, H., Ji, K., et al. 2020, ApJL, 900, L17
Xu, Z., Jiang, Y., Yang, J., et al. 2016, ApJL, 820, L21
Xu, Z., Jiang, Y., Yang, J., et al. 2017, ApJL, 840, L21
Yan, X. L., Liu, Z., Zhang, J., et al. 2020, ScChE, 63, 1656
Yurchyshyn, V., Kumar, P., Abramenko, V., et al. 2017, ApJ, 838, 32
Zhang, Y., Xu, Z., Zhang, Q., et al. 2022, ApJL, 933, L20
Zhang, Z., Chen, D.-Y., Wu, J., et al. 2019, RAA, 19, 160
Zuccarello, F., Guglielmino, S. L., Capparelli, V., et al. 2020, ApJ, 889, 65